\documentclass[aps,prl,preprint,showpacs,preprintnumbers,superscriptaddress]{revtex4-1}

\pdfoutput=1

\usepackage{graphicx}% Include figure files
\usepackage{xspace}

\RequirePackage{lineno}

\begin{document}

%\pagewiselinenumbers

\title{Indication of Electron Neutrino Appearance from an Accelerator-produced Off-axis Muon Neutrino Beam}

\newcommand{\superk}           {Super-Kamiokande\xspace}       
\newcommand{\nue}                {$\nu_{e}$\xspace}
\newcommand{\nuebar}           {$\bar{\nu}_{e}$\xspace}
\newcommand{\numubar}           {$\bar{\nu}_{\mu}$\xspace}
\newcommand{\numu}             {$\nu_{\mu}$\xspace}
\newcommand{\nutau}             {$\nu_{\tau}$\xspace}
\newcommand{\nux}                {$\nu_{x}$\xspace}
\newcommand{\numunue}       {$\nu_{\mu} \rightarrow \nu_{e}$\xspace}
\newcommand{\numunux}       {$\nu_{\mu} \rightarrow \nu_{x}$\xspace}
\newcommand{\numunutau}    {$\nu_{\mu} \rightarrow \nu_{\tau}$\xspace}

\newcommand{\tonethree}       {$\theta_{13}$\xspace}
\newcommand{\tonetwo}         {$\theta_{12}$\xspace}
\newcommand{\ttwothree}       {$\theta_{23}$\xspace}
\newcommand{\ssttmue}          {$\sin^2 2 \theta_{{\mu}e}$\xspace}
\newcommand{\sstonethree}    {$\sin^2 2 \theta_{13}$\xspace}
\newcommand{\ssttwothree}    {$\sin^2 2 \theta_{23}$\xspace}

\newcommand{\msqonetwo}   {$\Delta m^2_{12}$\xspace}
\newcommand{\msqonethree} {$\Delta m^2_{13}$\xspace}
\newcommand{\msqtwothree} {$\Delta m^2_{23}$\xspace}
\newcommand{\absmsqtwothree} {$|\Delta m^2_{23}|$\xspace}
\newcommand{\msqmue}        {$\Delta m^2_{{\mu}e}$\xspace}
\newcommand{\msqmumu}     {$\Delta m^2_{\mu\mu}$\xspace}

\newcommand{\enu}               {$E_{\nu}$\xspace}
\newcommand{\pmu}              {$p_{\mu}$\xspace}
\newcommand{\amome}         {$E_{e}$\xspace}
\newcommand{\evis}              {$E_{vis}$\xspace}
\newcommand{\pizero}           {$\pi^{0}$\xspace}
\newcommand{\pizerogg}       {$\pi^{0}\to\gamma\gamma$\xspace}

\newcommand{\degree}      {$^\circ$\xspace}

\newcommand{\INSTC}{\affiliation{University of Alberta, Centre for Particle Physics, Department of Physics, Edmonton, Alberta, Canada}}
\newcommand{\INSTDF}{\affiliation{The Andrzej Soltan Institute for Nuclear Studies, Warsaw, Poland}}
\newcommand{\INSTEE}{\affiliation{University of Bern, Albert Einstein Center for Fundamental Physics, Laboratory for High Energy Physics (LHEP), Bern, Switzerland}}
\newcommand{\INSTFE}{\affiliation{Boston University, Department of Physics, Boston, Massachusetts, U.S.A.}}
\newcommand{\INSTD}{\affiliation{University of British Columbia, Department of Physics and Astronomy, Vancouver, British Columbia, Canada}}
\newcommand{\INSTBR}{\affiliation{Brookhaven National Laboratory, Physics Department, Upton, New York, U.S.A.}}
\newcommand{\INSTGA}{\affiliation{University of California, Irvine, Department of Physics and Astronomy, Irvine, California, U.S.A.}}
\newcommand{\INSTI}{\affiliation{IRFU, CEA Saclay, Gif-sur-Yvette, France}}
\newcommand{\INSTCI}{\affiliation{Chonnam National University, Institute for Universe \& Elementary Particles, Gwangju, Korea}}
\newcommand{\INSTGB}{\affiliation{University of Colorado at Boulder, Department of Physics, Boulder, Colorado, U.S.A.}}
\newcommand{\INSTFG}{\affiliation{Colorado State University, Department of Physics, Fort Collins, Colorado, U.S.A.}}
\newcommand{\INSTCJ}{\affiliation{Dongshin University, Department of Physics, Naju, Korea}}
\newcommand{\INSTFH}{\affiliation{Duke University, Department of Physics, Durham, North Carolina, U.S.A.}}
\newcommand{\INSTBA}{\affiliation{Ecole Polytechnique, IN2P3-CNRS, Laboratoire Leprince-Ringuet, Palaiseau, France }}
\newcommand{\INSTEF}{\affiliation{ETH Zurich, Institute for Particle Physics, Zurich, Switzerland}}
\newcommand{\INSTEG}{\affiliation{University of Geneva, Section de Physique, DPNC, Geneva, Switzerland}}
\newcommand{\INSTDG}{\affiliation{H. Niewodniczanski Institute of Nuclear Physics PAN, Cracow, Poland}}
\newcommand{\INSTCB}{\affiliation{High Energy Accelerator Research Organization (KEK), Tsukuba, Ibaraki, Japan}}
\newcommand{\INSTED}{\affiliation{Institut de Fisica d'Altes Energies (IFAE), Bellaterra (Barcelona), Spain}}
\newcommand{\INSTEC}{\affiliation{IFIC (CSIC \& University of Valencia), Valencia, Spain}}
\newcommand{\INSTEI}{\affiliation{Imperial College London, Department of Physics, London, United Kingdom}}
\newcommand{\INSTGF}{\affiliation{INFN Sezione di Bari and Universit\`a e Politecnico di Bari, Dipartimento Interuniversitario di Fisica, Bari, Italy}}
\newcommand{\INSTNA}{\affiliation{INFN Sezione di Napoli and Universit\`a di Napoli, Dipartimento di Fisica, Napoli, Italy}}
\newcommand{\INSTPA}{\affiliation{INFN Sezione di Padova and Universit\`a di Padova, Dipartimento di Fisica, Padova, Italy}}
\newcommand{\INSTBD}{\affiliation{INFN Sezione di Roma and Universit\`a di Roma "La Sapienza", Roma, Italy}}
\newcommand{\INSTEB}{\affiliation{Institute for Nuclear Research of the Russian Academy of Sciences, Moscow, Russia}}
\newcommand{\INSTCC}{\affiliation{Kobe University, Kobe, Japan}}
\newcommand{\INSTCD}{\affiliation{Kyoto University, Department of Physics, Kyoto, Japan}}
\newcommand{\INSTEJ}{\affiliation{Lancaster University, Physics Department, Lancaster, United Kingdom}}
\newcommand{\INSTFC}{\affiliation{University of Liverpool, Department of Physics, Liverpool, United Kingdom}}
\newcommand{\INSTFI}{\affiliation{Louisiana State University, Department of Physics and Astronomy, Baton Rouge, Louisiana, U.S.A.}}
\newcommand{\INSTJ}{\affiliation{Universit\'e de Lyon, Universit\'e Claude Bernard Lyon 1, IPN Lyon (IN2P3), Villeurbanne, France}}
\newcommand{\INSTCE}{\affiliation{Miyagi University of Education, Department of Physics, Sendai, Japan}}
\newcommand{\INSTFJ}{\affiliation{State University of New York at Stony Brook, Department of Physics and Astronomy, Stony Brook, New York, U.S.A.}}
\newcommand{\INSTCF}{\affiliation{Osaka City University, Department of Physics, Osaka,  Japan}}
\newcommand{\INSTGG}{\affiliation{Oxford University, Department of Physics, Oxford, United Kingdom}}
\newcommand{\INSTBB}{\affiliation{UPMC, Universit\'e Paris Diderot, CNRS/IN2P3, Laboratoire de Physique Nucl\'eaire et de Hautes Energies (LPNHE), Paris, France}}
\newcommand{\INSTGC}{\affiliation{University of Pittsburgh, Department of Physics and Astronomy, Pittsburgh, Pennsylvania, U.S.A.}}
\newcommand{\INSTFA}{\affiliation{Queen Mary University of London, School of Physics, London, United Kingdom}}
\newcommand{\INSTE}{\affiliation{University of Regina, Physics Department, Regina, Saskatchewan, Canada}}
\newcommand{\INSTGD}{\affiliation{University of Rochester, Department of Physics and Astronomy, Rochester, New York, U.S.A.}}
\newcommand{\INSTBC}{\affiliation{RWTH Aachen University, III. Physikalisches Institut, Aachen, Germany}}
\newcommand{\INSTDD}{\affiliation{Seoul National University, Department of Physics and Astronomy, Seoul, Korea}}
\newcommand{\INSTFB}{\affiliation{University of Sheffield, Department of Physics and Astronomy, Sheffield, United Kingdom}}
\newcommand{\INSTDI}{\affiliation{University of Silesia, Institute of Physics, Katowice, Poland}}
\newcommand{\INSTDA}{\affiliation{STFC, Daresbury Laboratory, Warrington, United Kingdom}}
\newcommand{\INSTEH}{\affiliation{STFC, Rutherford Appleton Laboratory, Harwell Oxford, United Kingdom}}
\newcommand{\INSTCH}{\affiliation{University of Tokyo, Department of Physics, Tokyo, Japan}}
\newcommand{\INSTBJ}{\affiliation{University of Tokyo, Institute for Cosmic Ray Research, Kamioka Observatory, Kamioka, Japan}}
\newcommand{\INSTCG}{\affiliation{University of Tokyo, Institute for Cosmic Ray Research, Research Center for Cosmic Neutrinos, Kashiwa, Japan}}
\newcommand{\INSTF}{\affiliation{University of Toronto, Department of Physics, Toronto, Ontario, Canada}}
\newcommand{\INSTB}{\affiliation{TRIUMF, Vancouver, British Columbia, Canada}}
\newcommand{\INSTG}{\affiliation{University of Victoria, Department of Physics and Astronomy, Victoria, British Columbia, Canada}}
\newcommand{\INSTDJ}{\affiliation{University of Warsaw, Faculty of Physics, Warsaw, Poland}}
\newcommand{\INSTDH}{\affiliation{Warsaw University of Technology, Institute of Radioelectronics, Warsaw, Poland}}
\newcommand{\INSTFD}{\affiliation{University of Warwick, Department of Physics, Coventry, United Kingdom}}
\newcommand{\INSTGE}{\affiliation{University of Washington, Department of Physics, Seattle, Washington, U.S.A.}}
\newcommand{\INSTEA}{\affiliation{Wroclaw University, Faculty of Physics and Astronomy, Wroclaw, Poland}}
\newcommand{\INSTH}{\affiliation{York University, Department of Physics and Astronomy, Toronto, Ontario, Canada}}

\INSTC     %OK
\INSTDF    %OK
\INSTEE    %OK
\INSTFE    %OK 
\INSTD     %OK 
\INSTBR    %OK
\INSTGA    %OK
\INSTI     %OK
\INSTCI    %OK
\INSTGB    %OK
\INSTFG    %OK
\INSTCJ    %OK
\INSTFH    %OK
\INSTBA    %OK 
\INSTEF    %OK
\INSTEG    %OK
\INSTDG    %OK
\INSTCB    %OK
\INSTED    %OK 
\INSTEC    %OK
\INSTEI    %OK
\INSTGF    %OK
\INSTNA    %OK
\INSTPA    %OK
\INSTBD    %OK
\INSTEB    %OK
\INSTCC    %OK
\INSTCD    %OK
\INSTEJ    %OK
\INSTFC    %OK
\INSTFI    %OK
\INSTJ     
\INSTCE    %OK
\INSTFJ    %OK 
\INSTCF    %OK
\INSTGG    %OK
\INSTBB    %OK
\INSTGC    %OK
\INSTFA    %OK
\INSTE     %OK
\INSTGD    %OK
\INSTBC    %OK
\INSTDD    %OK
\INSTFB    %OK
\INSTDI    %OK
\INSTDA    %OK
\INSTEH    %OK 
\INSTCH    %OK
\INSTBJ    %OK
\INSTCG    %OK
\INSTF     %OK
\INSTB     %OK
\INSTG     %OK
\INSTDJ    %OK
\INSTDH    %OK
\INSTFD    %OK
\INSTGE    %OK
\INSTEA    %OK
\INSTH     %OK

\author{K.\,Abe}\INSTBJ
\author{N.\,Abgrall}\INSTEG
\author{Y.\,Ajima}\thanks{also at J-PARC Center}\INSTCB
\author{H.\,Aihara}\INSTCH
\author{J.B.\,Albert}\INSTFH
\author{C.\,Andreopoulos}\INSTEH
\author{B.\,Andrieu}\INSTBB
%\author{M.D.\,Anerella}\INSTBR
\author{S.\,Aoki}\INSTCC
\author{O.\,Araoka}\thanks{also at J-PARC Center}\INSTCB
\author{J.\,Argyriades}\INSTEG
% ...

\collaboration{The T2K Collaboration}\noaffiliation

\hfill\emph{Preprint submitted to PRL}

\date{\today}

\begin{abstract}
The T2K experiment observes indications of $\nu_\mu\rightarrow \nu_e$ appearance in
data accumulated with $1.43\times{}10^{20}$ protons on target.
Six events pass all selection criteria at the far detector.
In a three-flavor neutrino oscillation scenario with
$|\Delta m_{23}^2|=2.4\times 10^{-3}$~eV$^2$, $\sin^2 2\theta_{23}=1$ and $\sin^2 2\theta_{13}=0$, the 
expected number of such events is 1.5$\pm$0.3(syst.).
Under this hypothesis, the probability to observe six or more candidate events is 7$\times$10$^{-3}$,
equivalent to 2.5$\sigma$ significance.
At 90\%~C.L., the data are consistent with 0.03(0.04)$<\sin^2 2\theta_{13}<$ 0.28(0.34)  
for $\delta_{\rm CP}=0$ and normal (inverted) hierarchy. 
\end{abstract}


\maketitle

We report results of a search for $\nu_e$ appearance in the T2K experiment~\cite{Itow:2001ee}. 
In a three-neutrino mixing scenario, flavor oscillations are
described by the PMNS matrix~\cite{Maki:1962mu,Pontecorvo:1967fh}, usually parametrized
by the three angles $\theta_{12}$, $\theta_{23}$, $\theta_{13}$, and the {\it CP}-violating phase $\delta_{\rm CP}$.
Previous experiments have observed
neutrino oscillations driven by $\theta_{12}$ and $\theta_{23}$
in the solar (\msqonetwo) and atmospheric (\msqonethree$\simeq$\msqtwothree) sectors~\cite{ashie:2005ik,Hosaka:2005um,SNO05,Ahn:2006zza,KamLAND08,Adamson:2011ig}.
In the atmospheric sector, data are consistent with \absmsqtwothree$\simeq 2.4\times10^{-3}$~eV$^2$, a
normal \msqtwothree$>0$ or inverted \msqtwothree$<0$ mass hierarchy, and $\sin^22\theta_{23}$ close to, or equal to unity.  
Searches for oscillations driven by $\theta_{13}$
have been inconclusive and upper limits  have been derived~\cite{Yamamoto:2006ty,Wendell:2010md,SNO10,KamLAND11}, with the most stringent being
$\sin^2 2\theta_{13}$$<$0.15 (90\%C.L.), set by
CHOOZ~\cite{Chooz03} and MINOS~\cite{Adamson:2010uj}. 

T2K uses a conventional neutrino beam produced at J-PARC and
directed 2.5$^\circ$ off-axis to Super-Kamiokande (SK)
at a distance $L=295$~km.
This configuration produces a narrow-band \numu beam~\cite{beavis:bnl},  tuned at 
the first oscillation maximum $E_{\nu}=$\absmsqtwothree$L/(2\pi)\simeq$ 0.6~GeV, reducing backgrounds
from higher energy neutrino interactions.

Details of the T2K experimental setup are described
elsewhere~\cite{Abe:2011ks}. Here we briefly review the components relevant
for the $\nu_e$ search.
The J-PARC Main Ring (MR) accelerator~\cite{cite:Jparc} provides 30~GeV protons 
with a cycle of 0.3~Hz. 
Eight bunches are single-turn extracted 
in 5~$\mu$s and transported through an extraction line arc defined by superconducting combined-function magnets to the production target. 
The primary beamline is equipped with 21 electrostatic beam position monitors (ESM),
19 segmented secondary emission monitors (SSEM), one optical transition radiation monitor (OTR)
and five current transformers.
The secondary beamline, filled with He at atmospheric pressure,
is composed of the target, focusing horns and decay tunnel. The graphite target is 2.6~cm in diameter and 90~cm (1.9$\lambda_{int}$) long.
Charged particles exiting the target are sign 
selected and focused into the 96~m long decay
tunnel by three magnetic horns pulsed at 250~kA.  
Neutrinos are primarily produced in the decays of charged pions and kaons.
A beam dump is located at the end of the tunnel and is followed by muon monitors. 

The Near Detector complex~\cite{Abe:2011ks} located 280~m downstream from the target hosts
two detectors.
The on-axis Interactive Neutrino GRID (INGRID) accumulates neutrino interactions with high statistics to monitor the beam intensity,
direction and profile. It consists of 14 identical 7-ton iron-absorber/scintillator-tracker sandwich modules arranged in 10~m by 10~m crossed horizontal and vertical arrays centered on the beam.
The off-axis detector
reconstructs exclusive final states to study neutrino interactions
and beam properties corresponding to those expected at the far detector. 
Embedded  in the refurbished UA1 magnet (0.2~T), it consists of three 
large volume time projection chambers (TPCs)~\cite{Abgrall:2010hi} interleaved
with two fine-grained tracking detectors (FGDs, each 1~ton),
a $\pi^0$-optimized detector and a surrounding electromagnetic calorimeter.
The magnet yoke is instrumented as 
a side muon range detector.

The SK water Cherenkov far detector~\cite{fukuda:2002uc}
has a fiducial volume (FV) of  22.5~kton within its cylindrical inner detector (ID).
Enclosing the ID all around is the 2 m-wide outer detector (OD).
The front-end readout electronics allow for a 
zero-deadtime software trigger.
Spill timing information, synchronized by the Global Positioning System (GPS) 
with $<150$~ns precision,
is transferred online to SK and triggers the recording of
photomultiplier hits within $\pm$500 $\mu$s of the expected arrival time of the
neutrinos.

%\section{Event Selection} 
The results presented in this Letter are based on the first two 
physics runs: Run~1 (Jan--Jun 2010) and Run~2 (Nov 2010--Mar 2011).
During this time period, the MR proton beam power was continually increased and reached
145~kW with $9\times 10^{13}$ protons per pulse.
The targeting efficiency was monitored by the ESM, SSEM and OTR and found to be stable at over 99\%.
The muon monitors provided additional spill-by-spill steering information. 
A total of 2,474,419 spills were retained for analysis after beam and far detector quality cuts, yielding
$1.43\times10^{20}$ protons on target (p.o.t.).

We present the study of events in the far detector with only a single electron-like ($e$-like) ring.
The analysis produces a sample enhanced in \nue charged-current quasi-elastic interactions (CCQE) 
arising from $\nu_\mu\rightarrow \nu_e$ oscillations.
The main backgrounds are intrinsic \nue contamination in the beam
and neutral current (NC) interactions with a misidentified $\pi^0$.
The selection criteria for this analysis were fixed from Monte Carlo (MC) studies before the data 
were collected, optimized for the initial running conditions.
The observed number of events is compared to expectations based on 
neutrino flux and cross-section predictions for signal and all sources of backgrounds, which are
corrected using an inclusive $\nu_\mu$~charged-current (CC) measurement in the off-axis near detector.

\begin{figure}[tbp]
  \centering
 \includegraphics[width=0.75\textwidth]{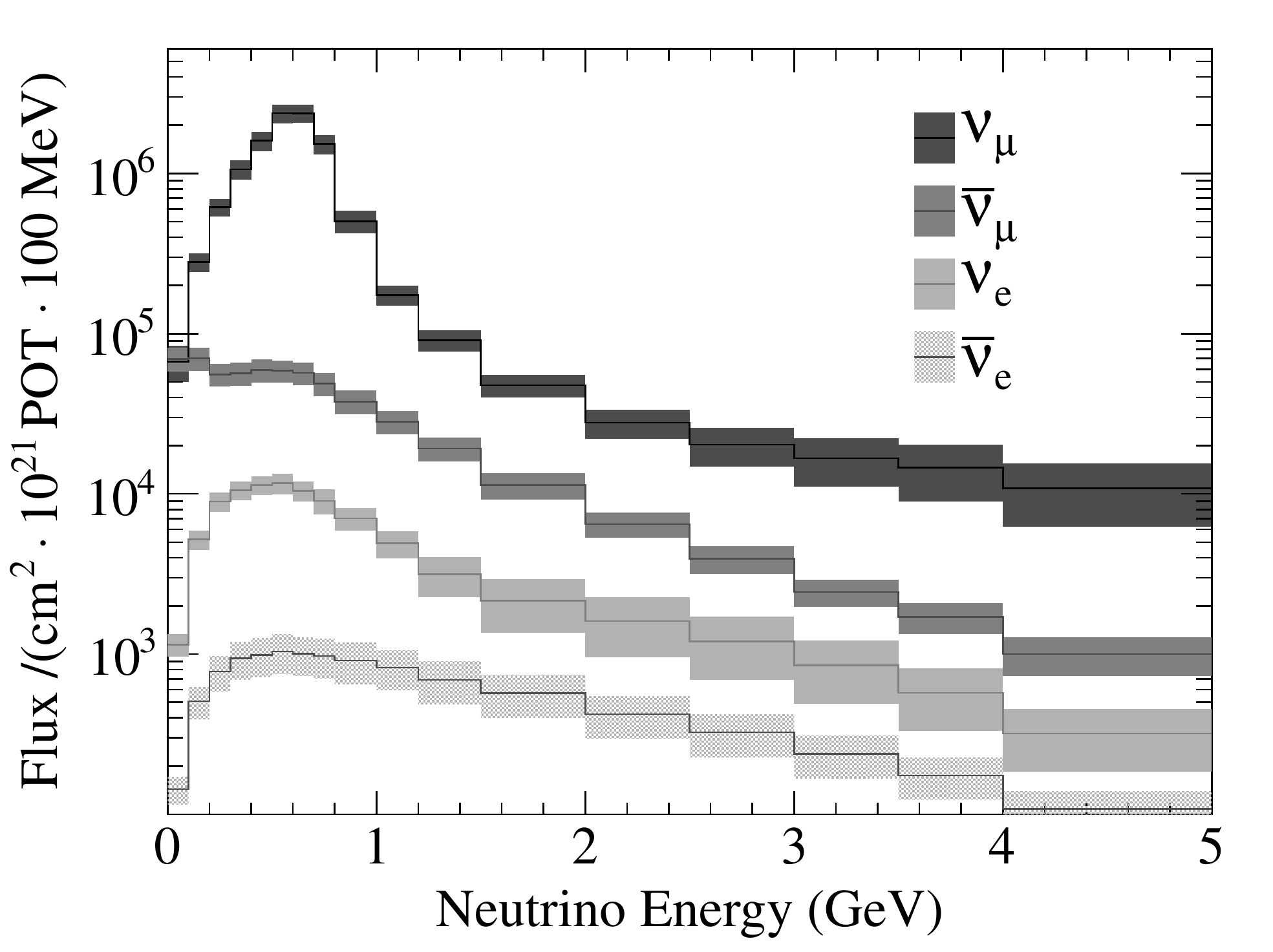}
 \caption{Predicted neutrino fluxes at the far detector, in absence of oscillations.
 The shaded boxes indicate the total
systematic uncertainties for each energy bin.}
 \label{fig:10d_flux}
\end{figure}

We compute the neutrino beam fluxes (Fig.~\ref{fig:10d_flux})  
starting from models and tuning them to experimental data.
Pion production in $(p,\theta)$ bins is based on the NA61 measurements~\cite{Abgrall:2011ae},
typically with 5--10\% uncertainties.
Pions produced outside the experimentally measured phase space, as well as kaons,
are modeled using FLUKA~\cite{cite:FLUKA1,cite:FLUKA2}. These pions are assigned systematic
uncertainties on their production of 50\%, while kaon production uncertainties, estimated from a comparison with data
from Eichten {\it et al.}~\cite{cite:Eichten}, range from 15\% to 100\% depending on the bin.  
GEANT3~\cite{cite:GEANT3}, with GCALOR~\cite{cite:GCALOR} for hadronic interactions,
handles particle propagation through the magnetic horns, target hall, decay 
volume and beam dump. 
Additional errors to the neutrino fluxes are included for the proton beam
uncertainties, secondary beamline component alignment uncertainties, and the
beam direction uncertainty.

\begin{figure}[!tbp]
  \centering
  \includegraphics[width=0.75\textwidth]{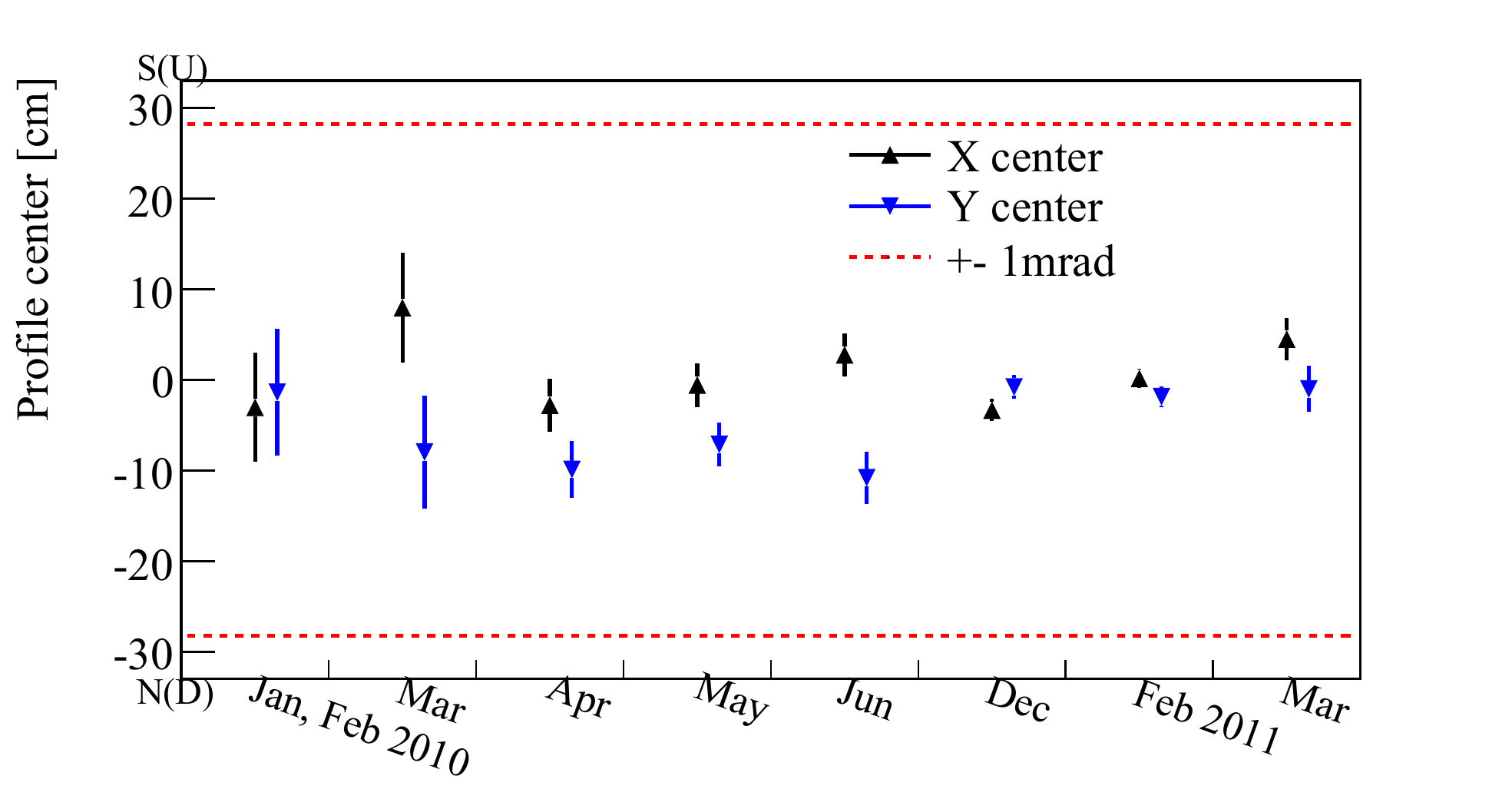}
  \caption{Beam centering stability in horizontal (x, South--North) and vertical (y, Down--Up) directions as a function
  of time, as measured by INGRID. Errors shown are only statistical.
  The dashed lines correspond to a change of beam direction by $\pm$1~mrad.}
  \label{fig:ingrideventsteer}
\end{figure}

The neutrino  beam profile and its absolute rate ($1.5$~events$/10^{14}$~p.o.t.) as measured by INGRID
were stable and consistent with expectations. 
The beam profile center (Fig.~\ref{fig:ingrideventsteer})
indicates that beam steering was better than $\pm 1$~mrad. 
The correlated systematic error  is
$\pm0.33(0.37)$~mrad for the horizontal(vertical) direction.
The error on the SK position relative to the beamline elements was obtained
from a dedicated GPS survey and is negligible.
As shown in Fig.~\ref{fig:10d_flux}, the estimated uncertainties of the intrinsic $\nu_\mu$ and $\nu_e$ fluxes below 1~GeV
are around 14\%.
Above 1~GeV, the intrinsic $\nu_e$ flux error is dominated by the uncertainty on the kaon
production rate with resulting errors of 20--50\%.

 \begin{table}[!tbp]
   \centering
   \caption{Summary of systematic uncertainties for the relative rate of
       different charged-current (CC) and neutral-current (NC) reactions
       to the rate for CCQE.}
   \begin{tabular}{lc}
     \hline
     \hline
   Process & Systematic error \\
       \hline
       CCQE & energy-dependent (7\% at 500 MeV) \\
       CC 1$\pi$ & 30\% ($E_\nu<2$~GeV) -- 20\% ($E_\nu>2$~GeV) \\
       CC coherent $\pi^\pm$ & 100\% (upper limit from~\cite{sciboone:cccoh}) \\
       CC other & 30\% ($E_\nu<2$~GeV) -- 25\% ($E_\nu>2$~GeV) \\
       NC 1$\pi^0$ & 30\% ($E_\nu<1$~GeV) -- 20\% ($E_\nu>1$~GeV) \\
       NC coherent $\pi$ & 30\% \\
       NC other $\pi$ & 30\% \\
       FSI & energy-dependent (10\% at 500 MeV) \\
         \hline
     \hline
   \end{tabular}
   \label{tab:systneut}
 \end{table}

The NEUT MC event generator~\cite{hayato:neut}, which has been
tuned with recent neutrino interaction data in an energy region
compatible with T2K~\cite{Nakajima:2010fp,AguilarArevalo:2010zc,Kurimoto:2009wq},
is used to simulate neutrino interactions in the near and far
detectors.
The GENIE~\cite{Andreopoulos:2009rq}
generator provides a separate cross-check of the assumed cross-sections and
uncertainties, and yields consistent results.  A list of
reactions and their uncertainties relative to the CCQE total
cross-section is shown in Table~\ref{tab:systneut}.   An energy-dependent error
on CCQE is assigned to account for the uncertainty in the low
energy cross-section, especially for the different
target materials between the near and far detectors.
Uncertainties in intranuclear final state interactions 
(FSI), implemented with a microscopic cascade model~\cite{salcedo:pionfsi},
introduce an additional error in the rates (see e.g. \cite{lee:pisummary}).

\begin{figure}[!tbp]
  \centering
  \includegraphics[width=0.75\textwidth]{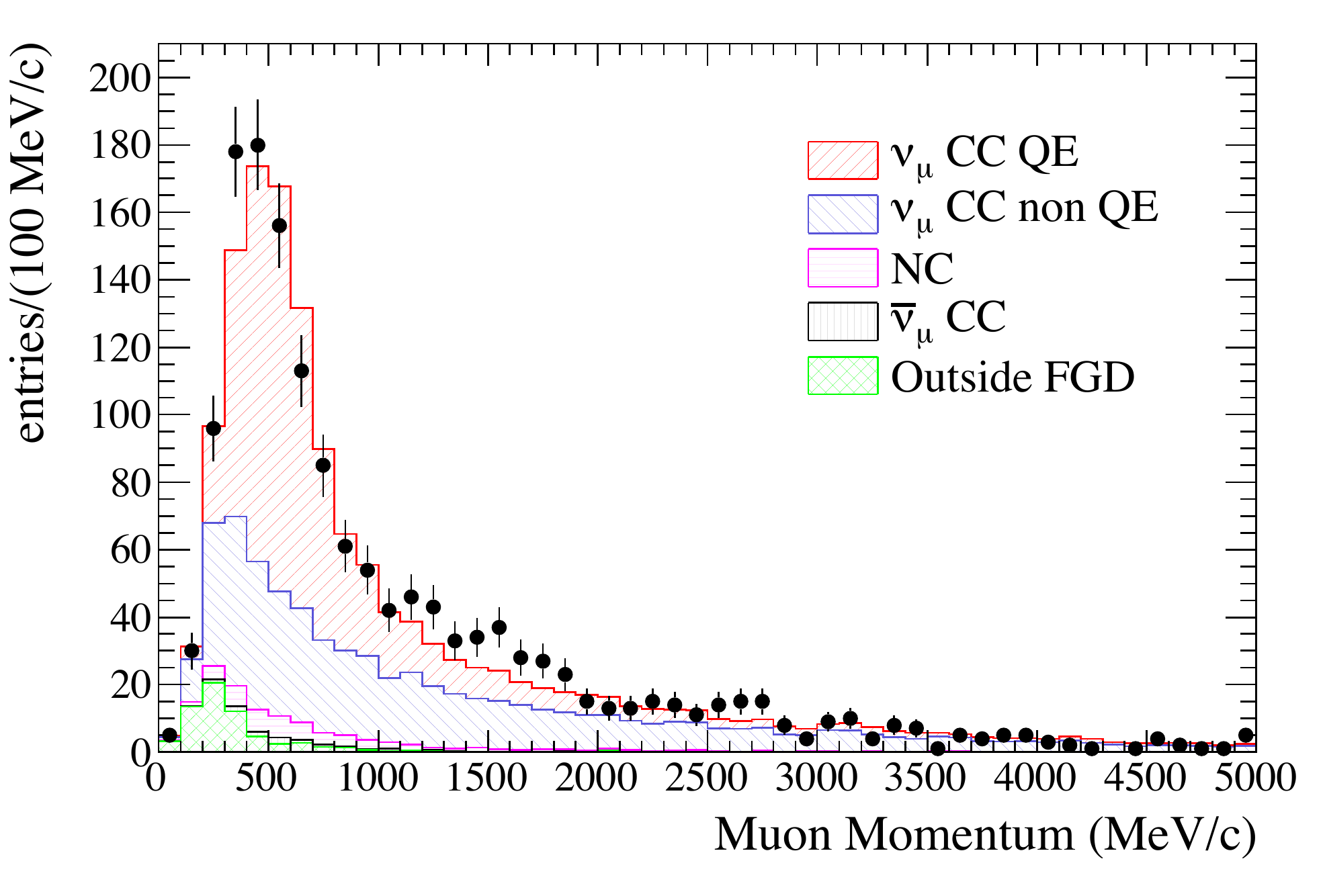}
 %% \includegraphics[width=2.9in]{MuMomentum10dv4.pdf}
%% \vspace{-0.5cm}
  \caption{Measured muon momentum of $\nu_\mu$~CC candidates reconstructed in the FGD target.
  The data are shown using points with error bars (statistical only) and the MC predictions are in 
  histograms shaded according to their type.
   }
  \label{fig:ND280momentum}
\end{figure}

An inclusive $\nu_\mu$~CC measurement in the off-axis near detector
is used to constrain the expected event rate at the far detector.
From a data sample collected in Run~1 and corresponding
to $2.88\times10^{19}$ p.o.t. after detector quality cuts,
neutrino interactions are selected in the FGDs with tracks entering the downstream TPC.
The most energetic negative track in the TPC is chosen and we
require its ionization loss to be compatible with a muon.
To reduce background from interactions outside the FGDs, there must be no track in the upstream TPC.
The analysis selects 1529 data events (38\% $\nu_\mu$~CC efficiency for 90\% purity, estimated from MC).
The momentum distribution of the selected muons (Fig.~\ref{fig:ND280momentum}) 
shows good agreement between data and MC. The measured data/MC ratio 
is
%\begin{eqnarray}
%R^{\mu,Data}_{ND}/R^{\mu,MC}_{ND}&=&1.036 \pm 0.028 (\mathrm{stat.}) ^{+0.044}_{-0.037} (\mathrm{det.syst.}) \nonumber\\
%& &\quad \quad \quad \quad \pm0.038(\mathrm{phys.syst.}),
%\label{eq:ratiodmc}
%\end{eqnarray}
\begin{equation}
R^{\mu,Data}_{ND}/R^{\mu,MC}_{ND} = 1.036 \pm 0.028 (\mathrm{stat.}) ^{+0.044}_{-0.037} (\mathrm{det.syst.})\pm0.038(\mathrm{phys.syst.}),
\label{eq:ratiodmc}
\end{equation}
where $R_{ND}^{\mu,Data}$ and $R_{ND}^{\mu,MC}$ are the p.o.t. normalized rates of $\nu_\mu$~CC interactions in data and MC.
The detector systematic errors mainly come from tracking and particle identification efficiencies,
and physics uncertainties are related to the interaction modeling.
Uncertainties that effectively cancel between near and far detectors were omitted.

\begin{figure}[!tbp]
   \centering
   \includegraphics[width=0.75\textwidth]{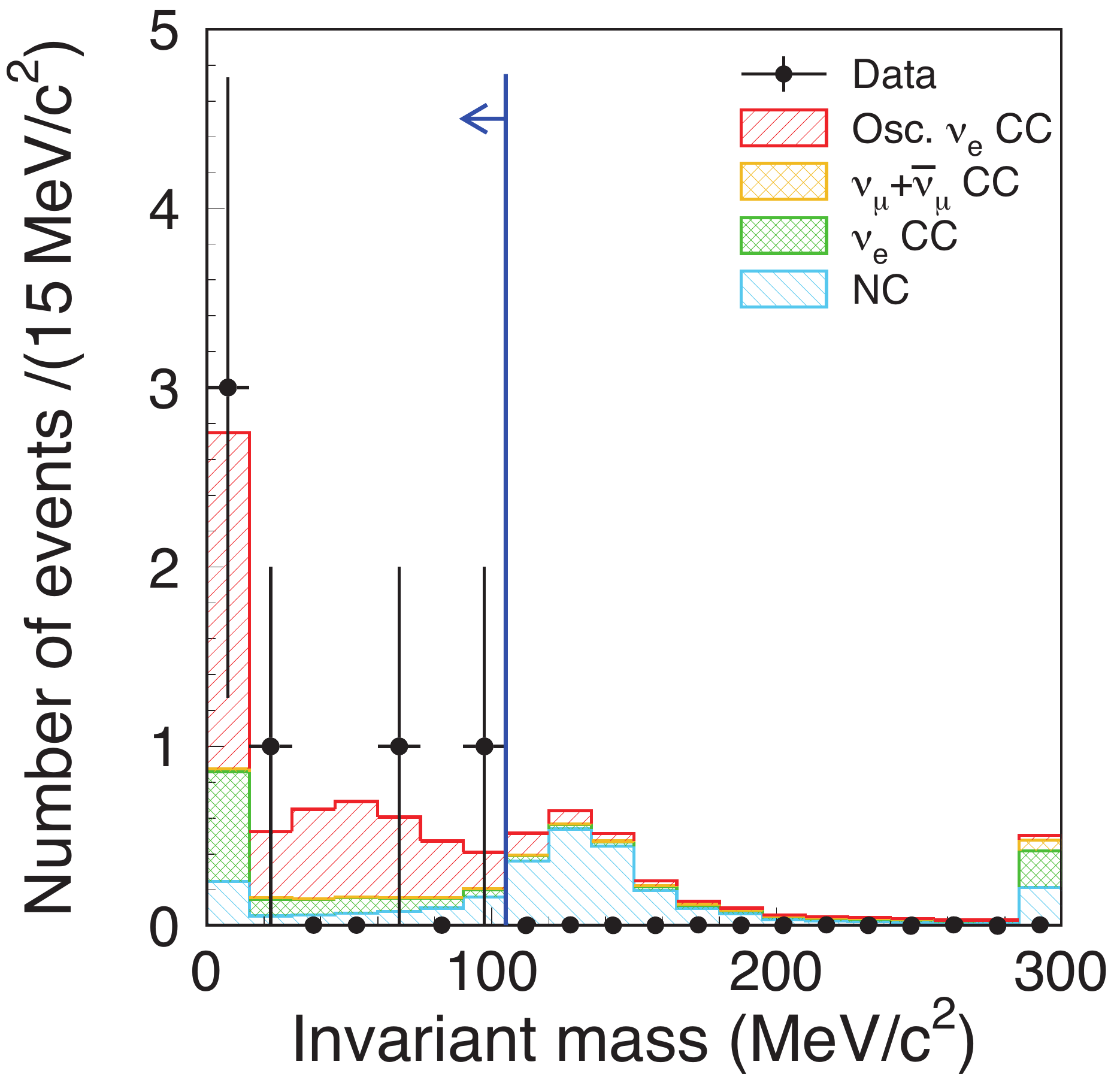}
   \caption{Distribution of invariant mass $M_{inv}$ when each event is forced to be reconstructed into two rings.
   The data are shown using points with error bars (statistical only) and the MC predictions are in 
  shaded histograms, corresponding to
  oscillated \nue CC signal 
   and various background sources for $\sin^22\theta_{13}=0.1$.
   The last bin shows overflow entries.
   The vertical line shows the applied cut at 105 MeV/$c^2$.
   }
   \label{fig:polfit}
 \end{figure}

At the far detector, we extract a fully-contained fiducial volume (FCFV) sample by requiring
no event activity in either the OD or 
in the 100 $\mu$s  before the event trigger time,
at least $30$ MeV electron-equivalent energy deposited in the ID (defined as visible energy $E_{vis}$), 
and the reconstructed vertex in the fiducial region.
The data have 88 such FCFV events that are within the timing range 
from $-$2 to 10~$\mu$s around the beam trigger time. 
The accidental contamination from non-beam related events
is determined from the sidebands to be $0.003$~events.
A Kolmogorov-Smirnov (KS) test of the observed number of FCFV events as a function of accumulated 
p.o.t. is compatible with the normalized event rate being constant ($p$-value=0.32).
The analysis relies on the well-established reconstruction techniques developed for 
other data samples~\cite{ashie:2005ik}.
Forty-one events are reconstructed with a single ring, and eight of those are $e$-like.
Six of these events have $E_{vis}>100$~MeV and
no delayed-electron signal.
To suppress misidentified $\pi^0$ mesons, the reconstruction of two
rings is forced by comparison of the observed and expected light patterns calculated 
under the assumption of two showers~\cite{Barszczak:2005sf}, and a cut
on the two-ring invariant mass $M_{inv}<105$~MeV$/c^2$ is imposed.
No events are rejected (Fig.~\ref{fig:polfit}).
\begin{figure}[!tbp]
  \centering
   \includegraphics[width=0.75\textwidth]{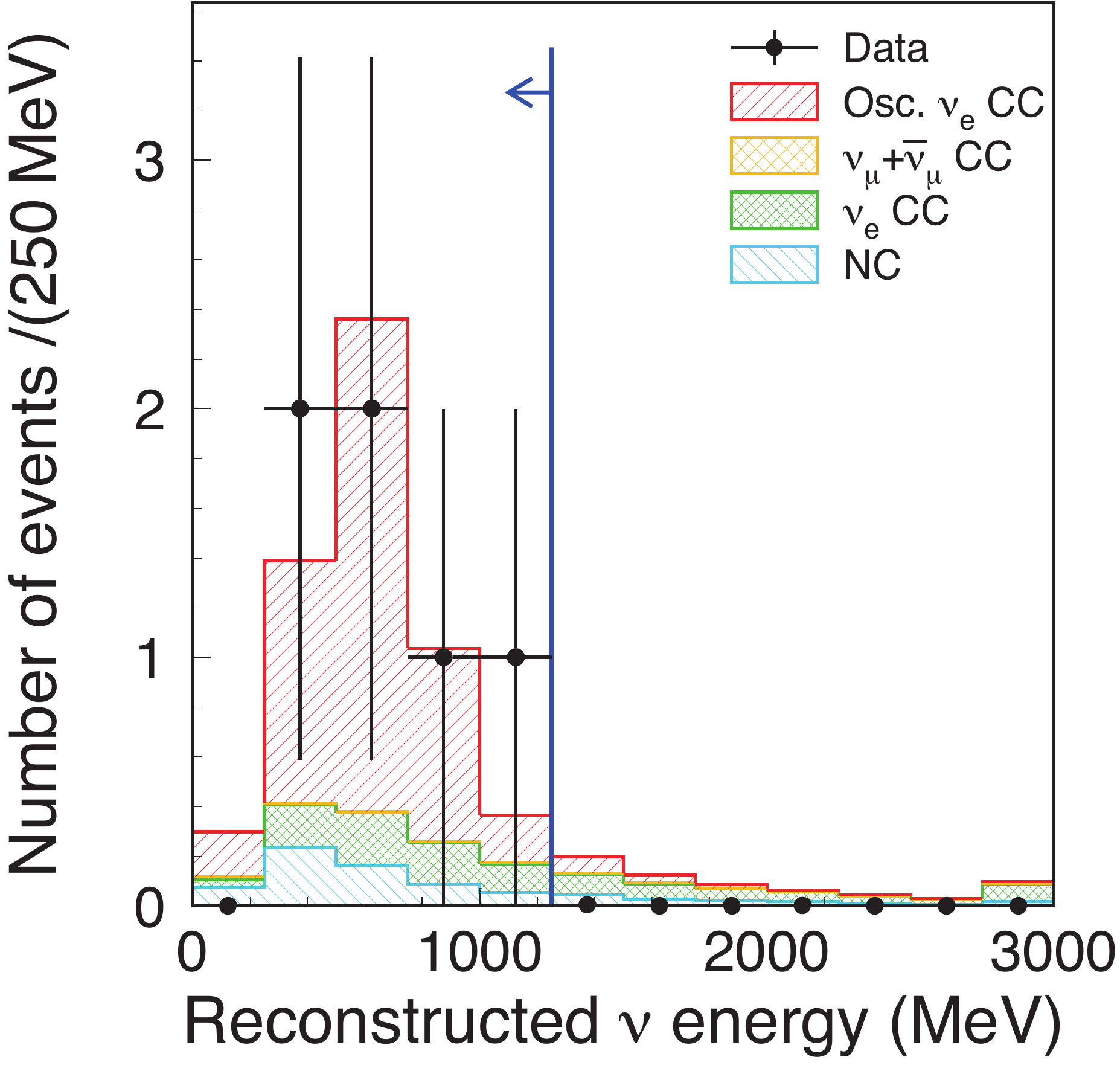}
  \caption{Same as Fig.~\ref{fig:polfit} for the
  reconstructed neutrino energy spectrum of the events
  which pass all \nue appearance signal selection criteria with the 
  exception of the energy cut.
  The vertical line shows the applied cut at 1250 MeV.
  }
  \label{fig:recoene}
\end{figure}
Finally, the neutrino energy $E^{rec}_{\nu}$ is computed using the reconstructed
momentum and direction of the ring, by assuming quasi-elastic kinematics
and neglecting Fermi motion.
No events are rejected by requiring $E^{rec}_{\nu}<1250$~MeV, aimed
at suppressing events from the intrinsic $\nu_e$ component arising
primarily from kaon decays (Fig.~\ref{fig:recoene}).
The data and MC reductions after each selection criterion are shown in Table~\ref{table:eventsummary}.
The $\nu_e$ appearance signal efficiency is estimated from MC to be 66\% while
rejection for $\nu_\mu+\bar\nu_\mu$ CC, intrinsic
\nue CC, and NC are $>99$\%, 77\%, and 99\%, respectively. 
Of the surviving background NC interactions constitute 46\%,
of which 74\% are due to $\pi^0$ mesons
and 6\% originate from single gamma production.
 \begin{table}[!tbp]
  \centering
  \caption{Event reduction for the \nue appearance search at the far detector.
   After each selection criterion is applied, the 
 numbers of observed (Data) and MC expected events of
 $\nu_\mu$~CC, 
 intrinsic \nue CC, 
 NC,  and the \nue CC signal, are given.
 All MC CC samples include three-flavor oscillations for $\sin^2 2\theta_{13}$=0.1 and $\delta_{\rm CP}=0$.
 }
  \begin{tabular}{lccccc}
   \hline
   \hline
                        & Data & \numu{}CC & \nue{}CC & NC &\numu$\rightarrow$\nue{}CC \\
   \hline
   (0)  interaction in FV   &  n/a    &       67.2    &      3.1   &    71.0    &       6.2 \\
   (1)~fully-contained FV & 88 & 52.4 & 2.9 & 18.3 & 6.0 \\
   (2)~single ring          & 41 & 30.8 & 1.8 & 5.7 & 5.2 \\
   (3)~$e$-like        &  8 &  1.0 & 1.8 & 3.7 & 5.2 \\
   (4)~$E_{vis}>100$~MeV   &  7 &  0.7 & 1.8 & 3.2 & 5.1 \\
   (5)~no delayed electron    &  6 &  0.1 & 1.5 & 2.8 & 4.6 \\
   (6)~non-$\pi^0$-like     &  6 &  0.04 & 1.1 & 0.8 & 4.2 \\
   (7)~$E^{rec}_{\nu}<1250$~MeV        &  6 &  0.03 & 0.7 & 0.6 & 4.1 \\
   \hline
   \hline
  \end{tabular}
  \label{table:eventsummary}
 \end{table}

Examination of the six data events shows properties consistent 
with $\nu_{e}$ CC interactions. 
The distribution of the cosine of the opening angle between the ring and the incoming beam direction
is consistent with CCQE events.
The event vertices in cylindrical coordinates ($R$,$\phi$,$z$)
show that these events are clustered at large $R$, 
near the edge of the FV in the upstream beam direction.
A KS test on the $R^2$ distribution of our final events yields a $p$-value of 0.03.
If this was related to contamination from penetrating particles produced in upstream neutrino
interactions,  then the ID region outside the FV should show evidence for such events, 
however this is not observed.
In addition, an analysis of the neutrino interactions occurring in the OD volume is consistent with expectations.

To compute the expected number of events at the far detector $N_{SK}^{exp}$,
we use the near detector $\nu_\mu$~CC interaction rate measurement as normalization, and 
the ratio of expected events in the near and far detectors, where common systematic errors cancel.
Using Eq.~\ref{eq:ratiodmc}, this can be expressed as:
\begin{equation}
N_{SK}^{exp} =\left({R_{ND}^{\mu,Data}}/{R_{ND}^{\mu,MC}} \right) \cdot N_{SK}^{MC},
\label{eq:nskexp}
\end{equation}
where $N_{SK}^{MC}$ is the MC number of events expected in the far detector.
Due to the correlation of systematic errors in the near and far detector samples, Eq.~\ref{eq:nskexp}
reduces the uncertainty on the expected number of events. 
Event rates are computed 
incorporating three-flavor oscillation probabilities and matter effects~\cite{PhysRevD.22.2718}
with $\Delta m_{12}^2=7.6\times 10^{-5}$~eV$^2$,
$\Delta m_{23}^2=+2.4\times 10^{-3}$~eV$^2$, $\sin^2 2\theta_{12} = 0.8704$,
$\sin^2 2\theta_{23} =1.0$, an average Earth
density $\rho$=3.2~g/cm$^3$ and $\delta_{\rm CP}=0$ unless otherwise noted.
The expectations  are
0.03(0.03)  \numu + \numubar CC, 0.8(0.7) intrinsic \nue CC, 
and 0.1(4.1) \numu $\rightarrow$ \nue oscillation events for $\sin^2 2\theta_{13}$=0(0.1),
and 0.6~NC events.
As shown in Table~\ref{table:syserr}, 
the total systematic uncertainty on $N_{SK}^{exp}$ depends on $\theta_{13}$.
Neutrino flux uncertainties contribute
14.9\%(15.4\%) to the far(near) event rates, but
their ratio has an 8.5\% error due to cancellations.
The near detector $\nu_\mu$~CC selection efficiency uncertainty yields $^{+5.6}_{-5.2}\%$ and
the statistical uncertainty gives 2.7\%.
The errors from cross-section modeling are dominated by FSI uncertainties
and by the knowledge of the $\sigma(\nu_e)/\sigma(\nu_\mu)$ ratio, estimated to $\pm 6\%$.
The systematic uncertainties due to event selection in SK
were studied with cosmic-ray muons, electrons from muon decays, and atmospheric neutrino events. 
Their contribution to $\delta N_{SK}^{exp}/N_{SK}^{exp}$ for e.g.  $\sin^22\theta_{13}=0.1$ is as follows:
1.4\% from the fiducial volume definition, 0.6\% from the energy scale and
0.2\% from the delayed electron signal tagging efficiency.
The $\pi^0$ rejection efficiency, studied with a NC $\pi^0$ topological control sample combining one
data electron and one simulated gamma event, contributes 0.9\%.
The uncertainty on the acceptance of one-ring $e$-like events 
was studied with an atmospheric neutrino sample, adding a contribution of 5\%
from ring counting and 4.9\% from particle identification uncertainties.
The performance of muon rejection by the ring particle identification algorithm
was investigated using cosmic-ray muons and atmospheric neutrino events, giving
0.3\%. The effect from uncertainties in the $M_{inv}$ cut is 6.0\%.
Combining the above uncertainties, the total
far detector systematic error contribution to $\delta N_{SK}^{exp}/N_{SK}^{exp}$ is 14.7\%(9.4\%)
for $\sin^22\theta_{13}=0(0.1)$. 

\begin{table}[btp]
\begin{center}
\caption{\small Contributions from various sources
 and the total relative uncertainty for $\sin^22\theta_{13}$=0 and 0.1, and $\delta_{\rm CP}=0$.}
\begin{tabular}{lrrr}
\hline \hline
         Source        & & ~~~$\sin^22\theta_{13}=0$  & ~~~$\sin^22\theta_{13}=0.1$      \\
\hline
(1)~neutrino flux        & &  $\pm$  8.5\%   & $\pm$ 8.5\%  \\
(2)~near detector    & &  ${}^{+5.6}_{-5.2}$\%  &  ${}^{+5.6}_{-5.2}$\%   \\
(3)~near det. statistics    & & $\pm$ 2.7\%    & $\pm$ 2.7\%  \\
(4)~cross section    & &   $\pm$ 14.0\%   & $\pm$ 10.5\%  \\
(5)~far detector     & &  $\pm$ 14.7\%    & $\pm$ 9.4\%  \\
\hline
Total  $\delta N_{SK}^{exp}/N_{SK}^{exp}$         & &  ${}^{+22.8}_{-22.7}$\%   &  ${}^{+17.6}_{-17.5}$\%  \\
\hline \hline
\end{tabular}
\label{table:syserr}
\end{center}
\end{table}

\begin{figure}[!tbp]
  \centering
  \includegraphics[width=0.75\textwidth]{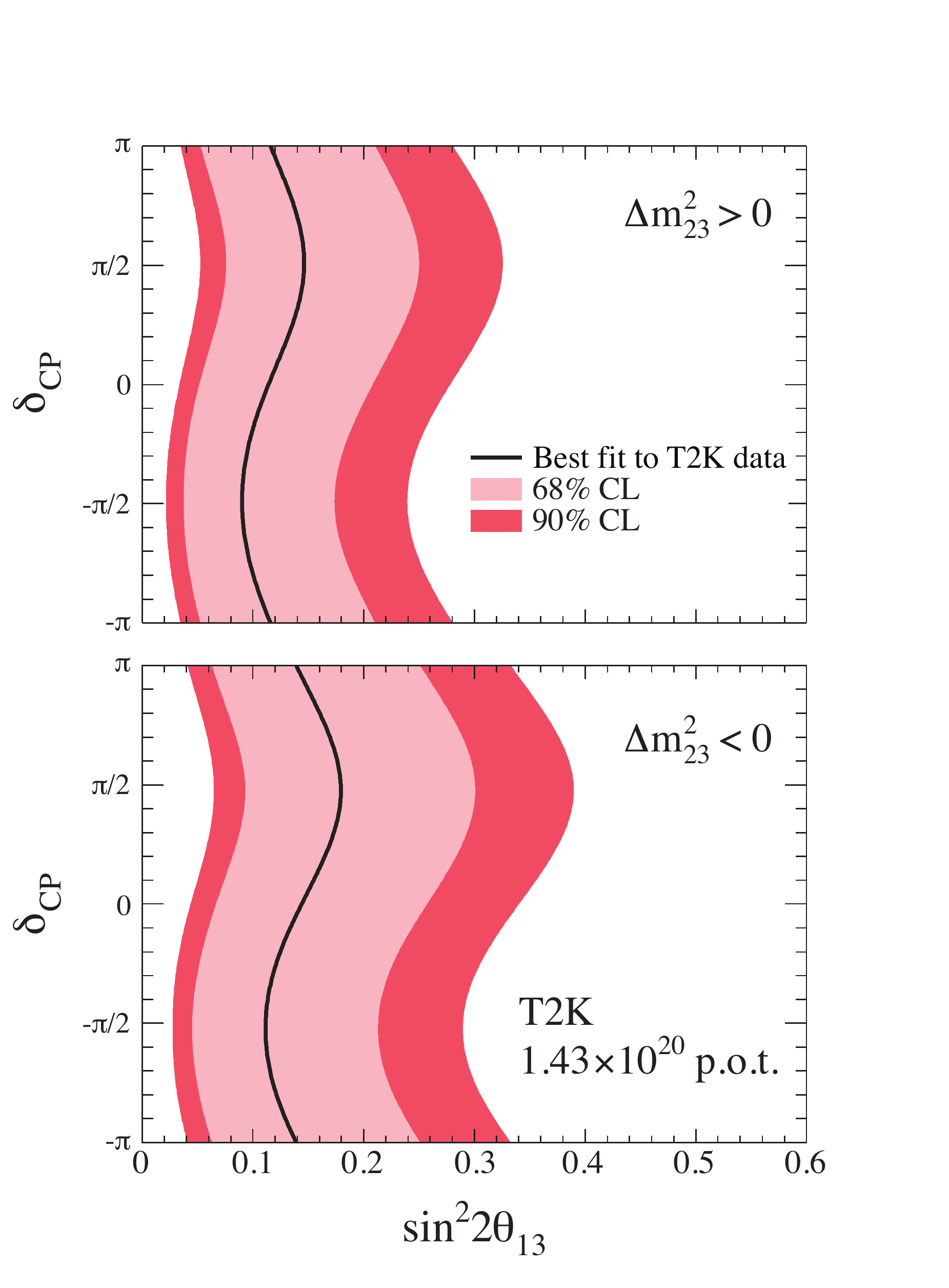}
 %% \includegraphics[height=4.2in]{result_dm2dcp_norm.pdf}
%%  \vspace{-0.4cm}
  \caption{
  %%%
  The 68\% and 90\%~C.L. regions for $\sin^{2}2\theta_{13}$ for each value of $\delta_{\rm CP}$,
           consistent with the observed number of events in the three-flavor oscillation case
           for normal (top) and inverted (bottom) mass hierarchy.
  %%%
  The other oscillation parameters are fixed (see text).
  The best fit values are shown with solid lines. 
           }
  \label{fig:normcontour}
\end{figure}

%\section{Results}
Our oscillation result is based entirely on comparing the number of $\nu_{e}$ candidate events with predictions,
varying $\sin^22\theta_{13}$ for each $\delta_{\rm CP}$ value.
Including systematic uncertainties, the expectation is 1.5$\pm$0.3(5.5$\pm$1.0)
events for $\sin^2 2\theta_{13}=0(0.1)$.
At each oscillation parameter point, a probability distribution for the expected number of events is 
constructed, incorporating systematic errors~\cite{cite:conradetal}, which is used to 
make the confidence interval (Fig.~\ref{fig:normcontour}), following the unified ordering prescription of 
Feldman and Cousins~\cite{cite:feldman_cousins}.

In conclusion, the observation of six single ring $e$-like events exceeds the expectation 
of a three-flavor neutrino oscillation scenario with $\sin^2 2\theta_{13} = 0$.
Under this hypothesis, the probability to observe six or more candidate events is 7$\times$10$^{-3}$.
Thus, we conclude that our data indicate $\nu_e$ appearance from a $\nu_\mu$ neutrino beam. 
This result converted into a confidence interval yields $0.03(0.04)<\sin^2 2\theta_{13}$ $<$ 0.28(0.34) at 90\%~C.L. for
$\sin^22\theta_{23}=1.0$,
$|\Delta m_{23}^2|$ = 2.4$\times$10$^{-3}$ eV$^2$, $\delta_{\rm CP}=0$ and
for normal (inverted) neutrino mass hierarchy. Under the same assumptions, the best fit points are 0.11(0.14), respectively.
For non-maximal $\sin^22\theta_{23}$, the confidence intervals remain unchanged to first order by replacing
$\sin^22\theta_{13}$ by $2\sin^2\theta_{23}\sin^22\theta_{13}$.
More data are required to firmly establish $\nu_e$ appearance and to better determine the angle $\theta_{13}$.

%\section{Acknowledgments}
\begin{acknowledgments}
We thank the J-PARC accelerator team for the superb accelerator performance and
CERN NA61 colleagues for providing essential particle production data and for their excellent collaboration.
We acknowledge the support of MEXT, Japan; 
NSERC, NRC and CFI, Canada;
CEA and CNRS/IN2P3, France;
DFG, Germany; 
INFN, Italy;
Ministry of Science and Higher Education, Poland; 
RAS, RFBR and the Ministry of Education and Science
of the Russian Federation; 
MEST and NRF, South Korea;
MICINN and CPAN, Spain;
SNSF and SER, Switzerland;
STFC, U.K.; and 
DOE, U.S.A.
We also thank CERN for their donation of the UA1/NOMAD magnet 
and DESY for the HERA-B magnet mover system.
In addition, participation of individual researchers
and institutions in T2K has been further supported by funds from: ERC (FP7), EU; JSPS, Japan; Royal Society, UK; 
DOE-OJI and DOE-Early Career program, and the A. P. Sloan Foundation, U.S.A.
\end{acknowledgments}

\bibliographystyle{apsrev4-1} 
\bibliography{references}

\end{document}